\documentstyle[emulateapj, times]{article}
\def\dgr{^\circ}
\newcommand{\etal}{{et al}\/.}

\def\subtitle{
  \vspace*{-6mm}
  \noindent
  {\scriptsize To appear in {\sc The Astrophyscal Journal}, 1998}}

\begin{document}
\accepted{8th April 1998}
\title{Extended and compact X-ray emission from the powerful radio galaxy 3C\,220.1}
\author{M.J.\ Hardcastle}
\affil{Department of Physics, University of Bristol, Tyndall Avenue,
Bristol BS8 1TL, UK}
\authoremail{M.Hardcastle@bristol.ac.uk}
\author{C.R.\ Lawrence}
\affil{Jet Propulsion Laboratory 169-506, Pasadena, California 91109, USA}
\authoremail{CRL@jplsp3.jpl.nasa.gov}

\and

\author{D.M.\ Worrall}
\affil{Department of Physics, University of Bristol, Tyndall Avenue,
Bristol BS8 1TL, UK\\and Harvard-Smithsonian Center for Astrophysics,
60 Garden Street, Cambridge, MA~02138, USA}
\authoremail{D.Worrall@bristol.ac.uk}
\begin{abstract}
We report on {\it ROSAT} HRI observations of the $z=0.61$ radio galaxy
3C\,220.1. The X-ray emission from this object consists of an extended
component, which we attribute to luminous cluster emission, and a
compact central source. The compact component is too bright to be
modelled as a cooling flow under some plausible assumptions for the
hot gas temperature and distribution; we suggest instead that it is
directly related to the core of the radio source. The X-ray flux of
the compact component is consistent with the prediction of Worrall
\etal\ (1994) that all powerful radio galaxies should have a central
jet-related X-ray emission component that is proportional in strength
to the radio core flux density. Other observations of
distant 3CR radio sources are consistent with this model.
\end{abstract}

 
\section{Introduction}

In unified models of powerful radio sources (e.g.\ Barthel 1989)
core-dominated quasars, lobe-dominated quasars and radio galaxies are
the same objects, with the apparent differences attributed to the
effects of relativistic beaming and anisotropic obscuration on a
population of sources oriented randomly with respect to the line of
sight. In quasars we see the AGN directly, while radio galaxies have
their jet axis at large angles to the line of sight so that a torus of
gas and dust obscures optical continuum and broad-line emission from
the AGN. This torus should also obscure soft X-ray emission
originating close to the central engines, leading to suggestions
(e.g.\ Crawford \& Fabian 1996) that X-ray emission from powerful
radio galaxies should in general be dominated by thermal emission from
hot cluster gas. However, an additional X-ray component may arise from
the radio-emitting plasma directly, through synchrotron or synchrotron
self-Compton radiation, or indirectly, through mechanisms such as the
inverse-Compton scattering of external photons. If the different
components can be separated, X-ray observations can provide physical
insights into the active nucleus, jets and large-scale environment.

There is substantial evidence that unabsorbed radio-related
non-thermal X-rays are seen. Fabbiano \etal\ (1984) first stressed
that a correlation between the total soft X-ray and radio-core
luminosity in radio galaxies implied a nuclear, jet-related origin for
at least some of the X-ray emission. More recent work has strengthened
this conclusion, as high-resolution X-ray observations have allowed
point-like and extended components to be separated (Worrall \&
Birkinshaw 1994; Edge \& R\"ottgering 1995; Worrall 1997).  Although
component separation is generally better for closer, less powerful
radio galaxies, compact soft X-ray emission is also seen in powerful
narrow-line FRII radio galaxies that are known to lie in sparse
environments, where cluster emission is not a source of confusion
(e.g.\ Hardcastle, Birkinshaw \& Worrall 1998a). The full complexity
is illustrated by X-ray observations of the nearby powerful cluster
radio galaxy Cygnus A. {\it EXOSAT} and {\it Ginga} have found
evidence for highly obscured core emission, ($N_H \sim 4 \times
10^{23}$ cm$^{-2}$; Arnaud \etal\ 1987, Ueno \etal\ 1994) which should
not have been seen with {\it ROSAT}, given its low-energy X-ray
passband. However, an unresolved core component was detected with the
{\it ROSAT} HRI (Harris, Perley \& Carilli 1994), implying that Cygnus
A's core has both an absorbed and an unabsorbed X-ray component; while
the absorbed component may be associated with emission from the AGN,
the unabsorbed component may be radio-related (Worrall
1997). Birkinshaw \& Worrall (1993) argued in the case of NGC 6251
that a plausible source for such compact X-ray emission is synchrotron
self-Compton (SSC) radiation from the base of the radio jet,
originating on scales larger than that of the torus and so avoiding
absorbtion. Such emission would be suppressed, but not eliminated, by
relativistic beaming effects in radio galaxies and [as discussed in
Worrall \etal\ (1994) and references therein] almost certainly
dominates the X-ray emission in core-dominated quasars.

It is therefore important in interpreting the X-ray emission from
radio galaxies to have spectral or spatial information
capable of distinguishing a non-thermal or compact component from
thermal or extended emission. High-redshift powerful radio galaxies
are important as counterparts to quasars in unified models, and
Worrall \etal\ (1994) performed such a spatial separation for the $z =
1$ radio galaxy 3C\,280. They showed that the unresolved
component fell on an extension of the correlation between X-ray and
radio core flux observed for core-dominated quasars, consistent with
the model discussed above. In this paper we report X-ray
observations of the radio galaxy 3C\,220.1 with the {\it ROSAT} HRI.
We use the high resolution of the HRI to constrain the contributions
from compact and extended emission.

3C\,220.1 is an FRII (Fanaroff \& Riley 1974) narrow emission-line
radio galaxy at $z=0.61$ (Spinrad \etal\ 1985). With $H_0 = 50$ km
s$^{-1}$ Mpc$^{-1}$ and $q_0 = 0$, used throughout the paper, its
178-MHz luminosity is $3.6 \times 10^{27}$ W Hz$^{-1}$
sr$^{-1}$. Radio imaging (Burns \etal\ 1984; Jenkins, Pooley \& Riley
1977; Harvanek \& Hardcastle 1998) shows it to be a typical
classical double object with largest angular size 35 arcsec (see Fig.\
\ref{contour}, inset); at this redshift one arcsecond corresponds to
8.94~kpc, so the projected linear size of 3C\,220.1 is about 300
kpc. Burns \etal\ (1984) report an unusually prominent one-sided jet
in the eastern lobe, and the radio core is also comparatively
prominent, which may be an indication that the source is significantly
affected by relativistic beaming, although no broad emission lines are
reported by Spinrad \etal\ (1985). The prominent radio core,
compared with the weak cores of the objects observed by Worrall \etal\
(1994), was the motivation for the present observations, since it allows us
to probe the possible core X-ray and radio association. Optical
observations show no evidence for a rich cluster near
3C\,220.1, although the presence of a gravitational lens arc with
$z=1.49$ in HST observations implies a deep potential well (M.\
Dickinson, private communication, 1997).

\section{Observations}

We observed 3C\,220.1 with the {\it ROSAT} HRI for a total of 36.2 ks
between 1995 Sep 12 and 1995 Sep 18. The data were analysed with the
IRAF Post-Reduction Off-line Software (PROS).

An X-ray source is detected with centroid at RA 09 32 39.4 DEC +79 06
27.2 (J2000 co-ordinates are used throughout). This is approximately 4
arcsec from the catalogued position of the optical identification of
3C\,220.1 (Laing, Riley \& Longair 1983) and from the position of the
radio core in the map of Burns \etal\ (1984), and we identify it with
the radio galaxy; offsets of up to 10 arcsec are consistent with the
absolute positional errors of {\it ROSAT}. There is only one
comparably bright source in the HRI field, at 09 31 20.6 +79 01 57.5
(350 arcsec from 3C\,220.1, which was on-axis); this is coincident
within 4 arcsec (in the same sense as for 3C\,220.1) with a faint,
blue, point source on sky survey plates (blue magnitude 18.72 in the
APS\footnote{The Automated Plate Scanner (APS) databases are supported
by the National Science Foundation, the National Aeronautics and Space
Administration, and the University of Minnesota, and are available at
URL: $<$http://aps.umn.edu/$>$.}  database) and a weak point-like
radio source [$15.5 \pm 0.6$ mJy at 1.4 GHz in the NRAO VLA Sky
Survey; Condon \etal\ (1997)], but is not identified with any known
object. For brevity, we refer to this source as `S2'. If the radio and
optical identifications of S2 are correct, the radio-optical spectral
index (0.4) and optical-X-ray spectral index (1.3) are consistent with
its being a background radio-loud AGN (e.g.\ Stocke \etal\ 1990).

\begin{table*}
\caption{Parameters of the $\beta$-models that best fit the radial
profile of 3C\,220.1}
\label{params}
\small
\begin{tabular}{rrrrrr}
$\beta$&Core radius&Central count density in $\beta$&Counts in point
model&$\chi^2$&Degrees of\\
&(arcsec)&model (counts arcsec$^{-2}$)&&&freedom\\
0.55&1.0&$18 \pm 2$&zero assumed&2.42&7\\
0.90&13.0&$0.352 \pm 0.065$&$117 \pm 20$&0.69&6\\
\end{tabular}
\end{table*}

Because we intended to search for extended emission, we defined a
large source region of radius 2 arcmin about the centroid of the X-ray
emission and a background annulus between 2 and 3 arcmin, excluding
regions around two weaker sources close to 3C\,220.1. In this region
the count rate for 3C\,220.1 is $(7 \pm 2) \times 10^{-3}$ counts
s$^{-1}$, or $260 \pm 70$ counts in total. With similar source and
background regions, S2's count rate is $(11 \pm 2) \times 10^{-3}$
counts s$^{-1}$. Vignetting reduces the 1-keV count rate by only $\sim
2$ per cent at the off-axis distance of S2 (David \etal\ 1997), so no
correction has been applied.

\begin{figure*}
\plotone{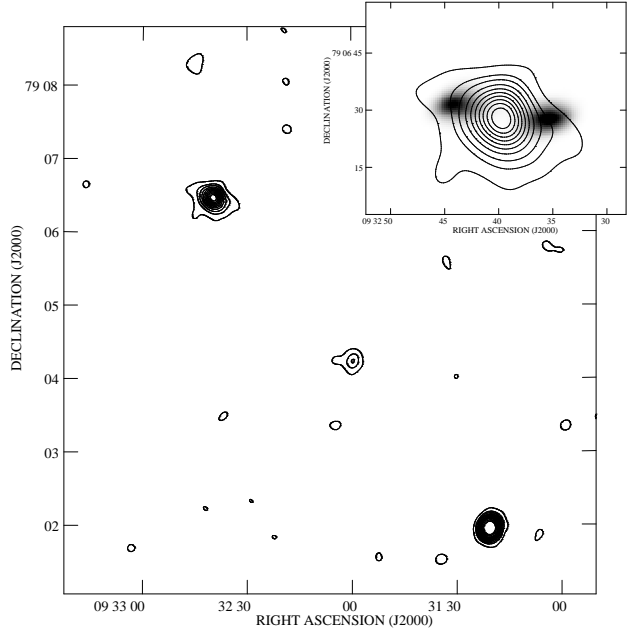}
\caption{X-ray image of the source associated with
3C\,220.1. 3C\,220.1 is in the NE corner, at the pointing centre of
the HRI; S2 is in the SW. The faint source between them is one of
those excluded from the spatial analysis. The HRI image has been
convolved with a Gaussian of $\sigma = 4$ arcsec. Contours are at
$0.0262, 0.045, 0.06, 0.08\dots 0.2, 0.25, 0.3, 0.5$ counts
pix$^{-1}$, where a pixel is 0.5 arcsec on a side. The lowest contour
is the 99.87 per cent confidence limit for Poisson noise at the
background level given this convolving Gaussian (Hardcastle, Worrall
\& Birkinshaw 1998b). Inset are the X-ray contours superposed on a
greyscale of the radio emission from 3C\,220.1 at 1.4-GHz (Harvanek \&
Hardcastle 1998) made with the B configuration of the
VLA. (Restoring beam $6.1 \times 3.9$ arcsec; black is 0.3 mJy
beam$^{-1}$.)}
\label{contour}
\end{figure*}

\begin{figure*}
\plottwo{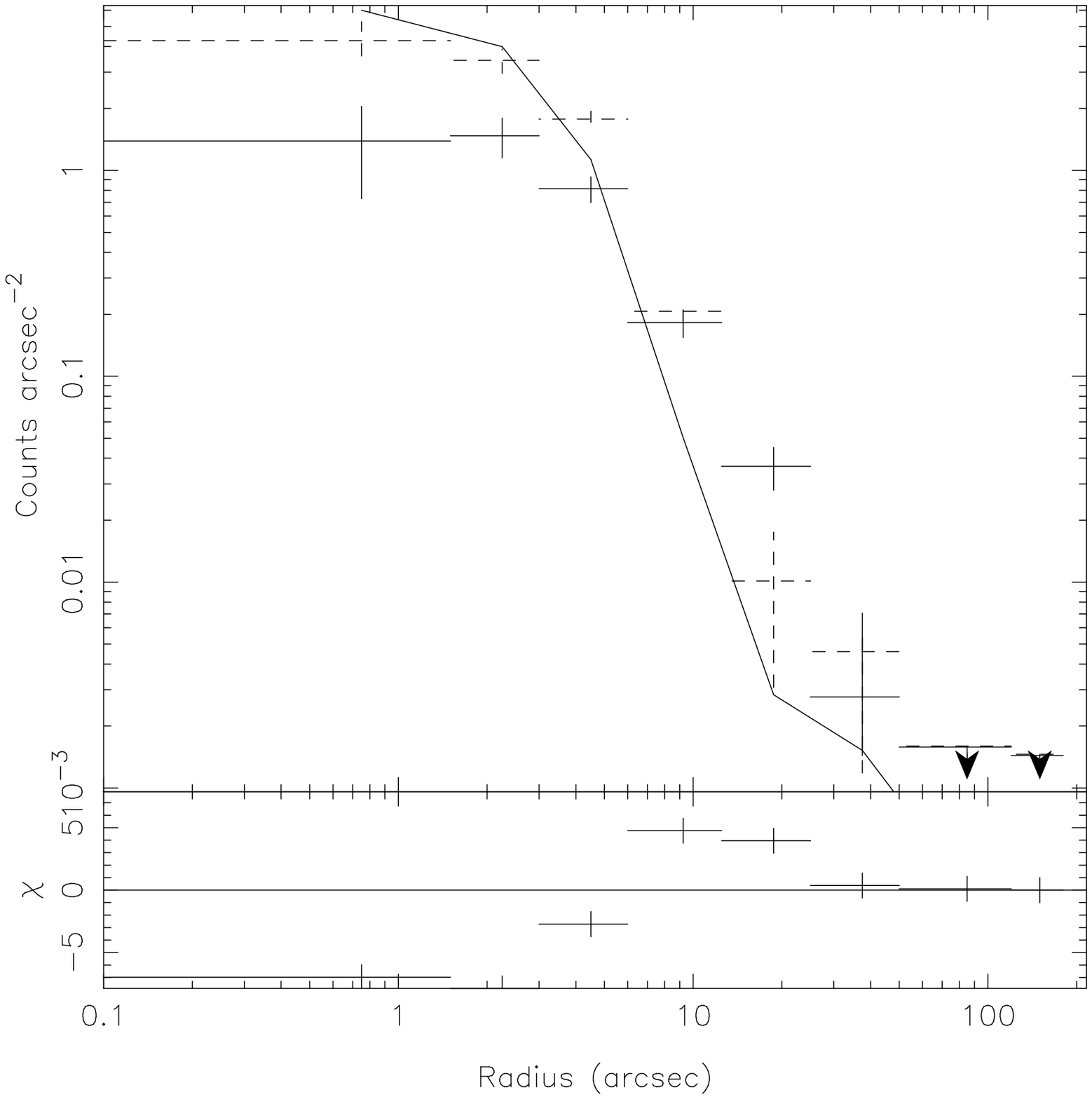}
{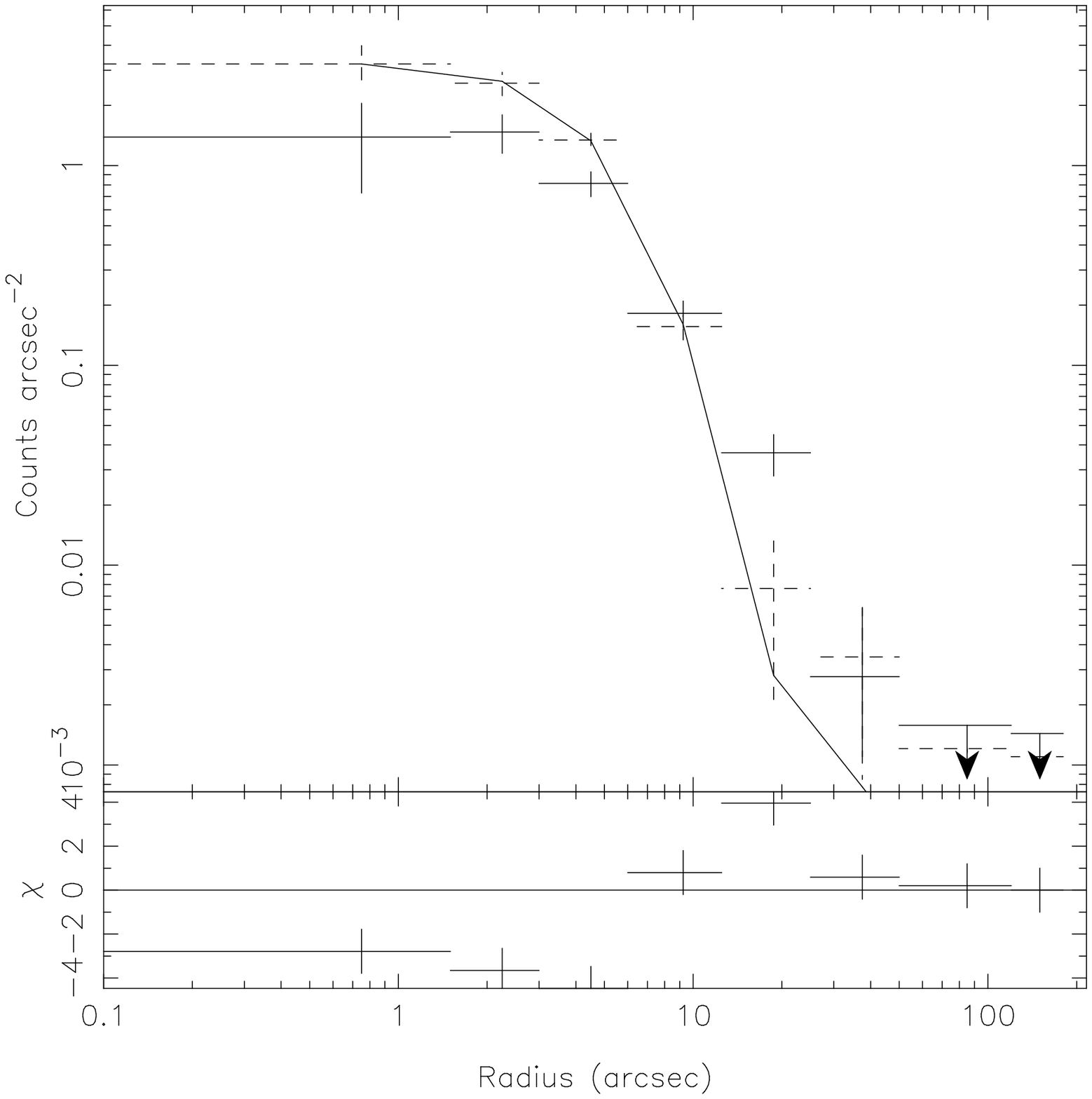}
\caption{Background-subtracted radial profiles of 3C\,220.1 (solid
crosses) and S2 (dashed crosses). On the left the nominal on-axis PSF
of the {\it ROSAT} HRI, integrated appropriately over the bins in use,
is plotted for comparison (solid line). On the right is plotted the
modified PSF selected as a good fit to the radial profile of S2,
similarly integrated. All profiles are normalised to the total counts
of 3C\,220.1. Below each figure are plotted the differences between
the PSF and 3C\,220.1's profile in each bin as a fraction of the error
in the bin. 3C\,220.1's profile can be seen to be a poor fit to either
PSF.}
\label{radial}
\end{figure*}

The source associated with 3C\,220.1 appears extended
(Fig. \ref{contour}), with an elongation in position angle $\sim
70\dgr$, close to the position angle of the radio galaxy axis
($79\dgr$). It is well known that spurious extension (`smearing') can
be produced by residual errors in the {\it ROSAT} aspect solution;
techniques for correcting this (e.g.\ Morse 1994) at present only work
on bright sources. However, any such smearing should apply equally to
all sources in a given field. In this case we see that the source S2,
comparable in brightness to 3C\,220.1, is less extended. Radial
profiles for 3C\,220.1 and S2 are compared with the nominal PSF for
the HRI in Fig.\ \ref{radial}. Since S2 is most likely to be a
point-like X-ray emitter (see above) the fact that it is by no means
perfectly fitted by the nominal HRI PSF suggests that there may be
some aspect smearing in the dataset. However, the radial profile of
3C\,220.1 is broader still, with a significant excess of counts on
scales $\sim 10$ arcsec from the centroid. This extension cannot be
produced by aspect smearing alone.

In order to separate point-like and extended components in 3C\,220.1
we need to estimate the form of the radially averaged PSF in the
presence of smearing. To do this we assume that S2 is a point
source. In general the radial profiles of off-axis sources like S2 are
broadened by the mirror blur of the XRT PSF. However, this effect is
not significant in the HRI at 6 arcmin off-axis; fits to the datasets
of short exposures of bright calibration objects provided with PROS
show that the effect becomes significant at $\ga 8$ arcmin, but that
objects at 6 arcmin are best-fitted with a parametrisation of the PSF
indistinguishable from the on-axis case. These results agree with the
analysis of David \etal\ (1997) and with their empirical expression
for the change in PSF with off-axis angle; G.\ Hasinger (private
communication, 1998) has analysed a large number of sources from the {\it
ROSAT} HRI Results Archive and come to the same conclusion.  We can
therefore conservatively estimate the effects of aspect smearing on
3C\,220.1 by using the radial profile of S2 as a description of the
PSF. This profile is well described ($\chi^2 = 2.1$ for 7 degrees of
freedom) by the nominal PSF broadened with a Gaussian of FWHM $\sim
4.5$ arcsec, whereas 3C\,220.1 is not acceptably fit with this model
($\chi^2 = 57$ for 7 d.o.f.; see Fig.~\ref{radial}.) In what follows we
use the broadened PSF based on S2 for model fitting.

The degree of extension in 3C\,220.1 may be quantified by fitting
models to the radial profiles of the sources. The models we use are
$\beta$-models (Sarazin 1986) or $\beta$-models with an additional
central point source. A $\beta$-model is physically appropriate if the
extended emission originates in hot gas in hydrostatic equilibrium and
[as discussed by Birkinshaw \& Worrall (1993)] provides a useful way
of assessing scales of extended emission even when the origins of the
emission are not known.  We choose $\beta$ from a range of possible
values between 0.35 and 0.9 and vary the core radius between 0.1 and
100 arcsec. The free parameters of the fit are the normalisation of
the $\beta$ model and of the point-like component if one is present;
there are therefore either 7 or 6 degrees of freedom in the fit.

\begin{figure*}
\epsscale{0.6}
\plotone{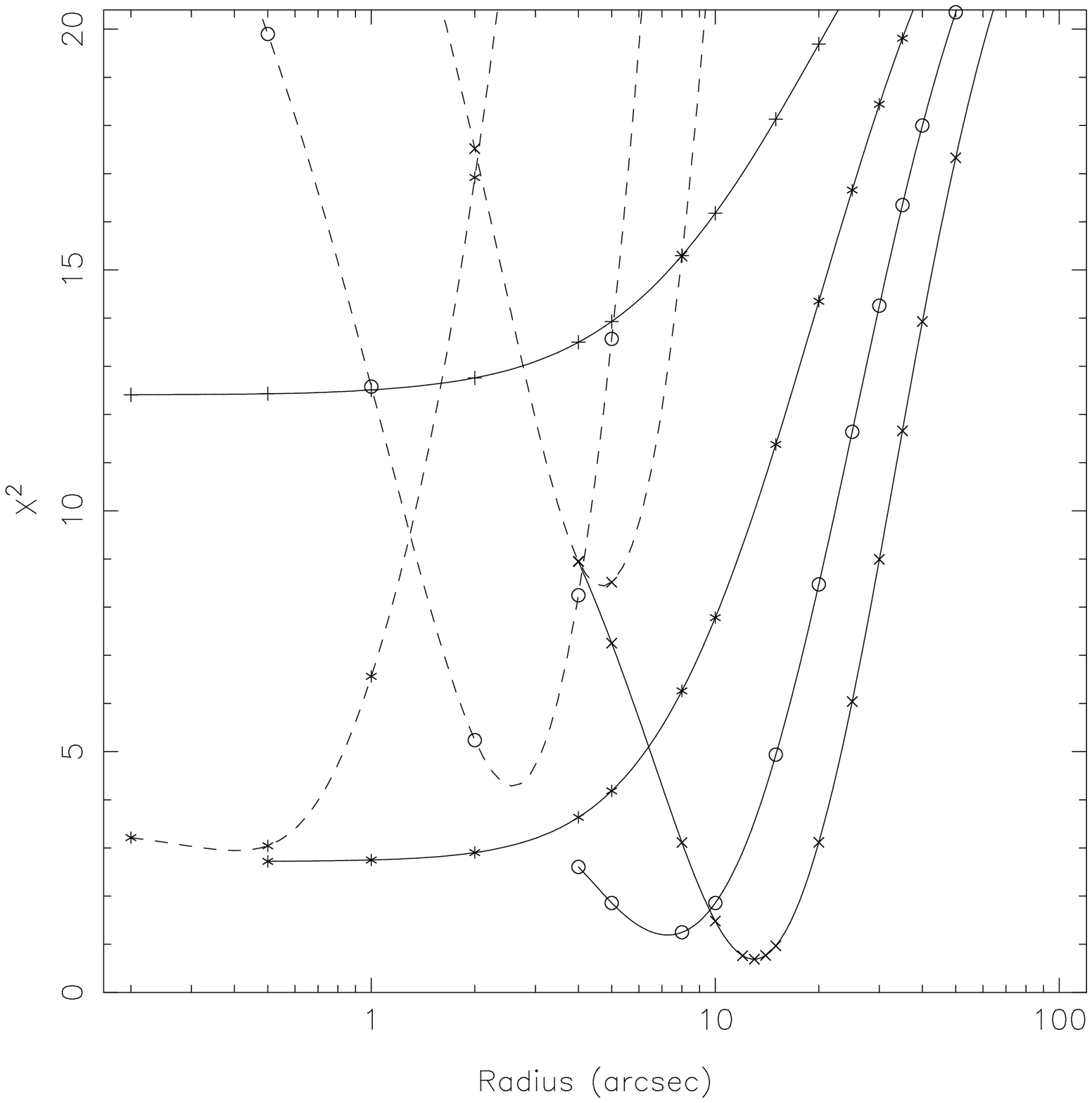}
\caption{$\chi^2$ as a function of model parameters for fits to the
radial profile of 3C\,220.1 using S2 as a `template' point
source. Points marked are the results of fits while the lines are
natural cubic splines drawn through the points. Solid lines show the
models with a point component (6 degrees of freedom), while dashed
lines show $\beta$-models only (7 degrees of freedom). Plusses
indicate fits with $\beta = 0.35$, stars $\beta=0.5$, circles $\beta =
0.67$ and Xs $\beta = 0.9$.}
\label{chi-modB}
\end{figure*}

The results obtained are shown in Fig.\ \ref{chi-modB}. The best-fit
model has $\beta = 0.9$ and a core radius of 13 arcsec ($\chi^2 = 0.7$
with 6 degrees of freedom). The point source in this model has a
count rate of $(3.2 \pm 0.6) \times 10^{-3}$ count s$^{-1}$, and so
provides 40 per cent of the total flux. The model with a point source
is significantly better (at the 99 per cent level on an F-test) than
the best-fit model consisting of a beta-model alone. Best-fit values
are tabulated in Table \ref{params}.

\section{Discussion}

\subsection{Origin of the extended emission}

The extended emission around 3C\,220.1 may be thermal radiation from
gas on cluster or group scales. We use the luminosity-temperature
relationship derived by David \etal\ (1993) to estimate a
self-consistent cluster luminosity and temperature, ignoring effects
of evolution. The steep best-fit $\beta$s mean that any additional
contribution to the luminosity from low-surface-brightness outer
regions below the detection threshold should be less than 15 per cent,
so errors caused by this effect are not significant. Using the
best-fit $\beta$ model the rest-frame 2-10 keV luminosity of the
extended emission would be $4.2 \times 10^{44}$ erg s$^{-1}$,
corresponding to a cluster temperature of 5.6 keV. [We assume galactic
absorption corresponding to $N_H = 1.9 \times 10^{20}$ cm$^{-2}$
(Stark \etal\ 1992) and use a Raymond-Smith model with elemental
abundance 30 per cent solar.] This luminosity is comparable to those
of other distant radio galaxies and to those of moderately rich
clusters [Crawford \& Fabian (1996); the 0.7-2 keV {\it total}
luminosity of the source, for comparison with their values, is $\sim
3.5 \times 10^{44}$ erg s$^{-1}$]. The core radius of 120 kpc is
comparable to those seen in nearby clusters of comparable luminosity
(e.g.\ Jones \& Forman 1984).

Birkinshaw \& Worrall (1993) provide convenient relationships between
$\beta$ and the normalising central count rate of a $\beta$-model and
physical parameters such as the density and pressure. Applying their
analysis to the best-fit $\beta$-model we find a central density of $1.6
\times 10^{-2}$ cm$^{-3}$ and a central cooling time for the gas
of 4.5 Gyr for $kT = 5.6$ keV. This corresponds to a cooling flow with mass
deposition rate $246 M_{\odot}$ yr$^{-1}$, although the average cooling
time within the core radius is comparable to the Hubble time. We
discuss the implications of a cooling-flow model below.

The fact that the X-ray emission has a similar apparent size to the
radio lobes and is extended roughly in the same direction might seem
to motivate an explanation in terms of inverse-Compton emission from
the lobes themselves, but investigation shows that neither the photon
density in the cosmic microwave background radiation nor that in a
putative hidden quasar (Brunetti, Setti \& Comastri 1997) is
sufficient to produce extended emission of this luminosity. We
therefore adopt a thermal model for the extended emission in what
follows.
\subsection{Origin of the point-like emission}
\label{cooling}

If the extended emission is thermal a cooling flow may be
present. Unresolved emission at this distance has a spatial scale of
$\la 50$ kpc, so without more detailed information on the spatial and
spectral X-ray properties of the source we cannot rule out the central
spike in a cooling flow as the origin of some or all
of the compact component of the X-ray emission. However, we can
estimate the contribution of a cooling flow by making some plausible
assumptions about the temperature and density variation of the gas.

We assume that the isothermal $\beta$-model with core radius $r_{\rm
core}$ and gas temperature $T_0$ (as estimated above) describes the
source adequately at large radii. Let $n_0$ be the central normalising
density for this $\beta$-model, as usual. We assume that cooling is
only important inside the cooling radius $r_{\rm cool}$; let the
density at the cooling radius be $n_{\rm cool}$. Then the density
profile in the cluster, assumed to be continuous, is taken to be
\[
n(r) = \cases{
n_{\rm cool} \left({r_{\rm cool}\over{r_{\rm inner}}}\right)
   &$r \le r_{\rm inner}$\cr
n_{\rm cool} \left({r_{\rm cool}\over{r}}\right)
   &$r_{\rm inner} \le r \le r_{\rm cool}$\cr
n_0\left(1+{r^2 \over r^2_{\rm core}}\right)^{-{3\over2}\beta}
   &$r_{\rm cool} \le r$\cr}
\]
$r_{inner}$ is a small inner limiting radius inside which density is
constant; this avoids an infinity at $r=0$. At $r_{\rm cool} \le r$,
the density follows the $\beta$ model; so $n_{\rm cool}$ is fixed, by
continuity, in terms of $n_0$, $r_{\rm core}$ and $r_{\rm
cool}$. Inside the cooling radius, the density is taken to vary as
$r^{-1}$. This dependence on radius is a reasonable approximation to
the correct, temperature-dependent behaviour (e.g.\ Sarazin
1986). Since cooling flows are slow, we assume that the pressure at a
given $r$ of the cooling gas is the same as the pressure in the
reference $\beta$-model at an equivalent radius. The temperature in
the cooling gas is then
\[
T(r) = T_0 {{n_0}\over n(r)} \left(1+{r^2 \over r^2_{\rm core}}
\right)^{-{3\over2}\beta} \] and this assumption also forces pressure
balance at $r = r_{\rm cool}$ and $r = r_{\rm inner}$. Finally, we use
PROS to compute the emissivity, in {\it ROSAT} HRI counts s$^{-1}$ per
unit emission measure, of the gas as a function of temperature, taking into
account the redshift of the source, galactic neutral hydrogen column
density, and the energy bandpass of the instrument. Models can then be
fitted to the radial profile of the source.

The choice of cooling radius is to some extent arbitrary. We used the
$\beta$-model fits to set $r_{\rm cool}$ as the radius where cooling
time was $<10^{10}$ yr; in the best-fit model this is 12 arcsec, but
the results are only weakly sensitive to moderate changes in cooling
radius. The model is also only weakly dependent on $r_{\rm inner}$; we
chose a value corresponding to 0.01 arcsec. The effect of the cooling
flow, as expected, is to reduce the contribution required from an
additional central point source, but with these assumptions the
cooling flow does not render a point source unnecessary. The best-fit
point source contributions in this model correspond to
$2.5_{-0.3}^{+0.5} \times 10^{-3}$ counts s$^{-1}$, so that
approximately 30 per cent of the central component is contributed by
the cooling flow; the errors assigned are $1\sigma$, derived by
allowing $\beta$ and core radius to vary and determining cooling
radius as above. As expected, the $\chi^2$ obtained in fitting the
cooling flow and point source model is indistinguishable from that due
to a $\beta$ model and point source.

Since the point-like component cannot, under these plausible
assumptions, all be attributed to a cooling flow, it may be more
directly associated with the AGN. Assuming a power-law spectrum with a
spectral index $\alpha = 0.8$, the counts associated with the compact
component (after subtraction of the cooling flow model) correspond to
a 1-keV flux density of $17_{-2}^{+3}$ nJy\footnote{For comparison,
when we performed exactly the same analysis using the nominal HRI PSF,
instead of that described by S2, we found a 1-keV flux of $6 \pm 5$
nJy.}, where the errors are the formal $1\sigma$ confidence limits
provided by the fitting process. We plot this flux against the 5-GHz
radio core flux density (25 mJy: Giovannini \etal\ 1988) together with
the objects from Fig.\ 3 of Worrall \etal\ (1994) in Fig.\
\ref{cores}. It will be seen that 3C\,220.1 has both a higher radio
core flux density and a higher core X-ray flux density than the object
(3C\,280) studied by Worrall \etal\ (1994); this is in the sense of
their prediction that X-ray core luminosity should correlate with
radio core luminosity. The X-ray flux density of the compact component
in 3C\,220.1 places it slightly above the line of slope unity plotted
by Worrall \etal\ through the core-dominated quasars. For comparison
we plot the X-ray and radio core flux density of Cygnus A, which can
be seen to lie very close to the line, in good agreement with the
correlation (Worrall 1997). Other existing data on high-redshift radio
galaxies (e.g. Crawford \& Fabian 1996) are also consistent with this
result, although unfortunately these additional sources are upper
limits in either or both of the X-ray (due to non-detection or
non-separation of unresolved emission) and radio (due to
undetected radio cores) flux densities.

\section{Conclusions}

\begin{figure*}
\plotone{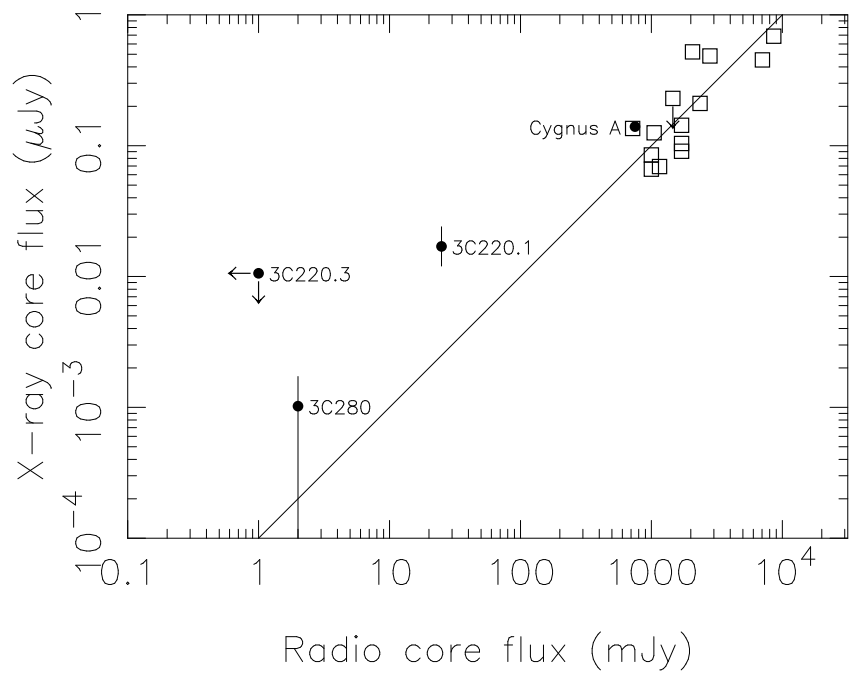}
\caption{Core X-ray and radio flux densities for 3C\,220.1, 3C\,220.3
and 3C\,280 together with core-dominated quasars matched in
isotropic radio power with the radio galaxies [adapted from Worrall
\etal\ (1994)] and Cygnus A [from Worrall (1997)]. A
line of slope unity is plotted for reference. Radio galaxies are
marked with filled circles and quasars with squares. Arrows indicate
upper limits.} 
\label{cores}
\end{figure*}

We observed 3C\,220.1 in order to test the hypothesis of Worrall
\etal\ (1994) that all powerful radio sources exhibit a compact X-ray
component related to their radio core. We find that there is strong
evidence for a compact X-ray component in this object, that it is too
bright to be attributed to a cooling flow under some simple
assumptions, and that it follows the expected positive correlation
between radio and X-ray core flux or luminosity. 3C\,220.1 lies
slightly above the line of slope unity plotted through the
core-dominated quasars (Fig.\ \ref{cores}). Together with its
unusually prominent radio core and one-sided radio jet, this may be an
indication that we are viewing the source at an angle to the line of
sight which is close to the radio galaxy-quasar boundary, so that its
central X-ray emission includes a non-jet-related component
originating close to the AGN.

Extended X-ray emission is also unequivocally detected around
3C\,220.1, with core radius and luminosity comparable to that of
nearby rich clusters. This is qualitatively consistent with the
HST detection of a gravitational lensing arc near the source.

\acknowledgements

We thank Mark Birkinshaw for providing the software used in radial
model fitting and for producing the cooling-flow model discussed in
section \ref{cooling}, and the referee, Dan Harris, for helpful
comments. This work was partially funded by NASA grant NAG 5-1882. MJH
acknowledges support from PPARC grant GR/K98582.  The Very Large Array
(VLA) is a facility of the National Radio Astronomy Observatory (NRAO)
which is operated by Associated Universities Inc., under cooperative
agreement with the National Science Foundation.

\noindent{\sc Note added in proof:}
The model discussed in section 3.2, while giving a good qualitative
representation of some cooling flows, is not a physically consistent
description of the cooling gas. (We thank A.C. Fabian for pointing
this out to us.) We have developed improved cooling-flow models in
which pressure varies as $r^{-1}$. These models exhibit a power-law
dependence of density and temperature on radius, roughly match the
mass implied by the gravitational arc, and have realistic mass
profiles. For reasonable choices of the model parameters, the central
X-ray profiles have broader wings than the model in section 3.2. When
fitted to the data, they tend to give fits worse than that provided by
the best-fit $\beta$ model with point source described in section
2. The limit on the contribution of a cooling flow to the X-ray flux
of the central source given in section 3 therefore remains valid. 


\begin{references}
\reference{24}Arnaud K.A., Johnstone R.M., Fabian A.C., Crawford C.S., Nulsen P.E.J., Shafer R.A., Mushotzky R.F., 1987, \mnras, 227, 241

\reference{35}Barthel P.D., 1989, \apj, 336, 606

\reference{71}Birkinshaw M., Worrall D.M., 1993, \apj, 412, 568

\reference{102}Brunetti G., Setti G., Comastri A., 1997, \aap, 325, 898

\reference{116}Burns J.O., Basart J.P., de Young D.S., Ghiglia D.C., 1984, \apj, 283, 515

\reference{144}Condon J.J., Cotton W.D., Greisen E.W., Yin Q.F., Perley R.A., Taylor G.B., Broderick J.J., 1997, available at URL $<$ftp://gibbon.cv.nrao.edu/pub/nvss/paper.ps$>$

\reference{161}Crawford C.D., Fabian A.C., 1996, \mnras, 282, 1483

\reference{167}David L.P., Harnden F.R., Kearns K.E., Zombeck M.V., Harris D.E., Prestwich A., Primini F.A., Silverman J.D., Snowden S.L., 1997, U.S. ROSAT Science Data Center report, available at URL: $<$http://hea-www.harvard.edu/rosat/rsdc\_www/hricalrep.html$>$

\reference{168}David L.P., Slyz A., Jones C., Forman W., Vrtilek S.D., 1993, \apj, 412, 479

\reference{184}Edge A.C., R\"ottgering H., 1995, \mnras, 277, 1580

\reference{195}Fabbiano G., Miller L., Trinchieri G., Longair M., Elvis M., 1984, \apj, 277, 115

\reference{200}Fanaroff B.L., Riley J.M., 1974, \mnras, 167, 31P

\reference{236}Giovannini G., Feretti L., Gregorini L., Parma P., 1988, \aap, 199, 73

\reference{259}Hardcastle M.J., Birkinshaw M., Worrall D.M., 1998a, \mnras, 294, 615

\reference{260}Hardcastle M.J., Worrall D.M., Birkinshaw M., 1998b,
\mnras, 296, 1098

\reference{99}Harvanek M., Hardcastle M.J., \apjs, in~press, astro-ph/9805363

\reference{267}Harris D.E., Perley R.A., Carilli C.L., 1994, in Courvoisier T.J.-L., Blecha A., eds, Multi-wavelength continuum emission of AGN, IAU Symposium 159, Kluwer, Dordrecht p.~375

\reference{310}Jenkins C.J., Pooley G.G., Riley J.M., 1977, \memras, 84, 61

\reference{314}Jones C., Forman W., 1984, \apj, 276, 38

\reference{360}Laing R.A., Riley J.M., Longair M.S., 1983, \mnras, 204, 151

\reference{443}Morse J.A., 1994, \pasp, 106, 675

\reference{556}Sarazin C.L., 1986, Rev.~Mod.~Phys., 58, 1

\reference{600}Spinrad H., Djorgovski S., Marr J., Aguilar L., 1985, \pasp, 97, 932

\reference{602}Stark A.A., Gammie C.F., Wilson R.W., Bally J., Linke R.A., Heiles C., Hurwitz M., 1992, \apjs, 79, 77

\reference{606}Stocke J.T., Morris S.L., Gioia I., Maccacaro T., Schild R.E., Wolter A., 1990, \apj, 348, 141

\reference{632}Ueno S., Koyama K., Nishida M., Yamauchi S., Ward M.J., 1994, \apj, 431, L1

\reference{667}Worrall D.M., 1997, in Ostrowski M., Sikora M., Madjeski G., Begelman M., eds, Relativistic jets in Active Galactic Nuclei, Astronomical Observatory of the Jagiellonian University, Cracow, p.~20 (astro-ph/9709165)

\reference{668}Worrall D.M., Birkinshaw M., 1994, \apj, 427, 134

\reference{671}Worrall D.M., Lawrence C.R., Pearson T.J., Readhead A.C.S., 1994, \apj, 420, L17

\end{references}
\end{document}